\documentclass[twocolumn,pra,nofootinbib,longbibliography]{revtex4-2}
\usepackage{amsmath,amsfonts,amssymb,amsthm}
\usepackage{graphicx,subfigure}
\usepackage{mathrsfs}
\usepackage[mathscr]{eucal}
\usepackage{mathtools}
\usepackage[T1]{fontenc}
\usepackage{physics}
\usepackage{bm}
\usepackage[bookmarks=false]{hyperref}
\usepackage{xcolor}
\usepackage{lineno}
\usepackage{url}
\usepackage{bbold}

\hypersetup{colorlinks=true,citecolor=blue,
linkcolor=blue,urlcolor=blue,pdfstartview=FitH,
bookmarksopen=true}

\newtheorem{thm}{Theorem}

\usepackage{pdfpages}
\usepackage{pgffor}
\makeatletter 
\AtBeginDocument{\let\LS@rot\@undefined} 
\makeatother

\newcommand{\Rrho}{\mathcal{R}_{\rho}}

\begin{document}
\title{Superiority of Krylov shadow tomography in estimating quantum Fisher information: From bounds to exactness}
\author{Yuan-Hao Wang}
\affiliation{Department of Physics, Shandong University, Jinan 250100, China}

\author{Da-Jian Zhang}
\email{zdj@sdu.edu.cn}
\affiliation{Department of Physics, Shandong University, Jinan 250100, China}

\begin{abstract}
Estimating the quantum Fisher information (QFI) is a crucial yet challenging task with widespread applications across quantum science and technologies. The recently proposed Krylov shadow tomography (KST) opens a new avenue for this task by introducing a series of Krylov bounds on the QFI. In this work, we address the practical applicability of the KST, unveiling that the Krylov bounds of low orders already enable efficient and accurate estimation of the QFI. We show that the Krylov bounds converge to the QFI exponentially fast with increasing order and can surpass the state-of-the-art polynomial lower bounds known to date. Moreover, we show that certain low-order Krylov bound can already match the QFI exactly for low-rank states prevalent in practical settings. Such exact match is beyond the reach of polynomial lower bounds proposed previously. These theoretical findings, solidified by extensive numerical simulations, demonstrate practical advantages over existing polynomial approaches, holding promise for fully unlocking the effectiveness of QFI-based applications.
\end{abstract}
\date{\today}
\maketitle

\noindent\textbf{\large Introduction}\\
The quantum Fisher information (QFI) is the cornerstone of quantum estimation theory \cite{1976Helstrom,2011Holevo}, describing the ultimate precision limit allowed by quantum mechanics \cite{1994Braunstein3439,2017Seveso12111,2020Zhang23418}. Ever since its inception in $1967$ \cite{1967Helstrom101}, the QFI has played a key role in quantum metrology and, more broadly, in quantum information science \cite{Toth2014JPAMT}, with enormous applications ranging from parameter estimation \cite{Zhang2015PRL,2015Lu7282,2018Zhou78,2019Bai40402,Xu2020PRL,Yang2020nQI,Zhang2022nQI,Ullah2023PRR,Zhou2024PRR,Zhang2024PRL,Zhou2025PRA,Montenegro2025PR} and entanglement detection \cite{Boixo2008PRL,Pezze2009PRL,Toth2018PRL,Toth2020PRL,Ren2021PRL,Tan2021PRL,Yang2024PRL} to variational quantum algorithms \cite{Meyer2021Q,Jiao2023AQT} and quantum machine learning \cite{Haug2024PRL}.

The widespread applications of the QFI underscore the imperative need for its efficient estimation, which is a crucial yet challenging task for large-scale quantum systems \cite{Gebhart2023NRP}. A fundamental difficulty is that the QFI is a highly nonlinear functional that cannot be expressed as a polynomial of the density matrix. This aligns with the general intuition that physical quantities of greater functional complexity are inherently more
difficult to estimate. {To circumvent this difficulty, numerous methods have been proposed for efficient estimation of the QFI \cite{Apellaniz2017PRA,Cerezo2021QST,Gacon2021simultaneous,Rath2021PRL,Vitale2024PQ,Halla2025estimationofquantum}}.
A common strategy in the literature is to alternately estimate some polynomial lower bounds on the QFI rather
than the QFI itself. Notable examples include the lower bound based on the Legendre transform \cite{Apellaniz2017PRA}, the lower bound based on the sub-QFI \cite{Cerezo2021QST}, and the family of lower bounds derived from the Taylor expansions \cite{Rath2021PRL,Vitale2024PQ}. These bounds, being functionals of lower complexity than the QFI, are more readily accessible in experiments. However, the efficiency and accuracy of these polynomial approaches are intrinsically limited. The underlying reason is that no polynomial lower bound can match the QFI exactly for all states. The resultant gap introduces unavoidable systematic errors in QFI estimation. Crucially, unlike statistical errors stemming from finite measurements, systematic errors cannot be mitigated by simply increasing experimental repetitions. The elimination of the gap is therefore of paramount importance for achieving reliable estimates of the QFI but, unfortunately, is unattainable with the polynomial approaches.

In this work, we address this challenge by further developing our recent proposal, called Krylov shadow tomography (KST) \cite{Zhang2025PRL}, which marks a paradigm shift from polynomial to non-polynomial approaches. The KST is an integration of the Krylov subspace method---a celebrated tool from applied mathematics \cite{Higham2015}---into the framework of shadow tomography \cite{Huang2020,Elben2022NRP}. The innovation of the KST is to significantly reduce and even close the gap by introducing a strict hierarchy of non-polynomial lower bounds on the QFI, referred to as the Krylov bounds. These bounds can approximate the QFI increasingly better and eventually match it exactly after a finite number of increments in order.
While the recent study \cite{Zhang2025PRL} laid the foundational groundwork for the KST, several key questions remain open: (i) How rapidly do the Krylov bounds converge to the QFI in general settings? (ii) How do their tightness and convergence compare with existing polynomial lower
bounds? (iii) Under what condition can the Krylov bounds match the QFI exactly at low orders? 
These questions are central to the practical applicability of the KST, given the fact that both
the number of measurements and the cost of postprocessing in shadow tomography \cite{Huang2020,Elben2022NRP} scale exponentially
with the order of the Krylov bounds {(see Ref.~\cite{Zhang2025PRL} for detailed description)}. 

The purpose of this work is to answer these questions. We prove that the Krylov bounds converge to the QFI exponentially fast with increasing order and surpass the state-of-the-art polynomial lower bounds known to date \cite{Rath2021PRL,Vitale2024PQ}. Moreover, we show
that certain low-order Krylov bounds are capable of matching the QFI exactly whenever the
state under consideration is of low rank. In conjunction with the fact that low-rank states arise frequently in numerous quantum information processing tasks \cite{Gross2010,Liu2012}, this exact match offers an intriguing pathway to yield accurate estimates of the QFI
with low resource demands. Furthermore, we demonstrate all these analytical findings through numerical simulations. This work therefore establishes the KST as a practically superior framework for QFI estimation, potentially advancing the experimental implementations of
QFI-based applications.

{We clarify that the present work is a valuable and nontrivial extension of our prior study \cite{Zhang2025PRL}. The new contribution here includes an exponential convergence theorem, a strict comparison with the most powerful polynomial lower bounds, and an explicit characterization of the exact match. These results put the KST on a much firmer footing and provide crucial insights into its practical applicability, which were not addressed in our previous work. }

\begin{figure}
    \centering
    \includegraphics[width=\linewidth]{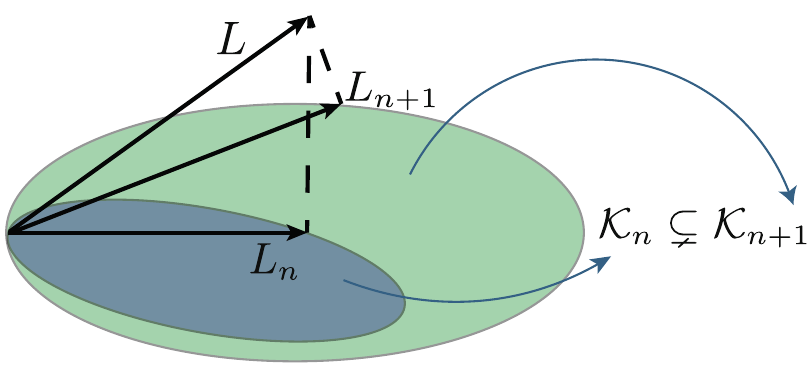}
    \caption{\textbf{Basic idea of the Krylov shadow tomography.} We construct a nested sequence of Krylov subspaces $\mathcal{K}_n$ and associate each of them with the Krylov bound defined by $B_n^{(\mathsf{Kry})}=\norm{L_n}_\rho^2$, where $L_n\in\mathcal{K}_n$ is chosen to be as close as possible to the SLD, represented by the symbol $L$ in the plot. As $\mathcal{K}_n$ continuously expands, $B_n^{(\mathsf{Kry})}$ approximates $F_Q$ increasingly better and eventually matches it exactly when $n=n^*$.
    }
    \label{fig1}
\end{figure}

\vspace{2em}
\noindent\textbf{\large Results}\\
\textbf{Basics of the Krylov shadow tomography}\\
To present our results clearly, we need to recapitulate some basics of the KST \cite{Zhang2025PRL}. Throughout this work, all operators are assumed to be Hermitian unless specified otherwise. Let $\rho$ be the unknown state under consideration. The QFI of $\rho$ with respect to an operator $H$ is defined as
\begin{eqnarray}
    F_Q=2\sum_{p_k+p_l>0}\frac{(p_k-p_l)^2}{p_k+p_l}\abs{\bra{k}H\ket{l}}^2,
\end{eqnarray}
where $p_k$ and $\ket{k}$ are the eigenvalues and eigenvectors of $\rho$, respectively.
We introduce the space of Hermitian operators
\begin{eqnarray}
    \mathcal{X}\coloneqq\{X: X_{kl}=0~~\text{if}~~p_k=p_l=0\},
\end{eqnarray}
where $X_{kl}=\bra{k}X\ket{l}$. That is,
the potentially nonzero elements of $X\in\mathcal{X}$ are only $X_{kl}$'s with $p_k+p_l>0$. 
For example, when $\rho=\begin{pmatrix}1 & 0 \\ 0 & 0 \end{pmatrix}$, all elements in $\mathcal{X}$ are of the form 
\begin{equation}
    X=\begin{pmatrix}* & * \\ * & 0 \end{pmatrix},
\end{equation}
where $*$ denotes the possibly nonzero entries of $X$.
We endow the space $\mathcal{X}$ with the following weighted inner product,
\begin{eqnarray}\label{weighted-inner-product}
    \langle X, Y\rangle_\rho\coloneqq\tr\left[\rho(XY+YX)/2\right],
\end{eqnarray}
for $X, Y\in\mathcal{X}$. The basic idea of the KST is to construct a nested sequence of subspaces of $\mathcal{X}$---called the Krylov subspaces---and optimally approximate $F_Q$ within each Krylov subspace.
The Krylov subspace of order $n$ is defined as
\begin{equation}
    \mathcal{K}_n \coloneqq \operatorname{span} \left\{i[\rho, H], \Rrho(i[\rho, H]), \cdots,\Rrho^{n-1}(i[\rho, H])\right\},
\end{equation}
that is, $\mathcal{K}_n$ is a subspace of $\mathcal{X}$ defined by linearly spanning the set $\{ \Rrho^k(i[\rho, H]) \}_{k=0}^{n-1}$. Here $\mathcal{R}_\rho$ denotes the superoperator
\begin{eqnarray}
    \mathcal{R}_\rho(X)\coloneqq\frac{1}{2}\left(\rho X+X\rho\right)
\end{eqnarray}
and $\mathcal{R}_\rho^k$ represents applying $\mathcal{R}_\rho$ iteratively $k$ times.
Apparently, $\mathcal{K}_{n+1}$ includes $\mathcal{K}_{n}$ as a subset, implying that $\mathcal{K}_n$ expands as $n$ increases. This expansion, however, must eventually terminate at a certain integer $n^*$, since the dimension of any $\mathcal{K}_n$ cannot exceed that of $\mathcal{X}$. We thus have
\begin{eqnarray}
    \mathcal{K}_1\subsetneq\mathcal{K}_2\subsetneq\cdots\subsetneq\mathcal{K}_{n^*}=\mathcal{K}_{n^*+1}=\mathcal{K}_{n^*+2}=\cdots.
\end{eqnarray}
Associated to each $\mathcal{K}_n$ for $n\leq n^*$, the Krylov bound is given by
\begin{equation}
    B_n^{(\mathsf{Kry})}\coloneqq\norm{L_n}_\rho^2.
\end{equation}
Here $\norm{X}_\rho\coloneqq\sqrt{\langle X, X\rangle_\rho}$ denotes the norm induced by the inner product \eqref{weighted-inner-product}. $L_n$ is an operator in $\mathcal{K}_n$ that is chosen to be as close as possible to the symmetric logarithmic derivative (SLD) (see Fig.~\ref{fig1}) \cite{2009Paris125}. The main finding is \cite{Zhang2025PRL}
\begin{eqnarray}\label{main-finding-1st}
    B_1^{(\mathsf{Kry})}<B_2^{(\mathsf{Kry})}<\cdots<B_{n^*}^{(\mathsf{Kry})}=F_Q,
\end{eqnarray}
stating that $B_n^{(\mathsf{Kry})}$ is an increasingly better lower bound on $F_Q$ as $n$ increases and eventually matches $F_Q$ exactly when $n=n^*$. Moreover, each $B_n^{(\mathsf{Kry})}$, albeit being non-polynomial in nature,  can be expressed in terms of some polynomials of $\rho$ and hence can be estimated via shadow tomography \cite{Zhang2025PRL}. The KST therefore enables systematically improvable and ultimately accurate estimation of the QFI.

\vspace{1em}
\noindent\textbf{Efficient QFI estimation with low-order Krylov bounds}\\
We clarify that the number of measurements and the cost of postprocessing required to estimate $B_n^{(\mathsf{Kry})}$ scale exponentially with the order $n$  \cite{Zhang2025PRL}. This overhead arises because evaluating $B_n^{(\mathsf{Kry})}$ requires estimating polynomials of $\rho$ up to degree $2n+1$  \cite{Zhang2025PRL}. It is well established that shadow tomography
incurs exponential cost when used to estimate such polynomials \cite{Huang2020,Elben2022NRP}. That is, the exponential overhead is a generic feature
of shadow tomography \cite{Zhou2024nQI,Peng2025PRA} rather than a peculiarity of our
KST. Due to this overhead, we are restricted to efficiently accessing only
low-order Krylov bounds in practice. Consequently, it is crucial to figure out how tightly these
low-order bounds approximate the QFI. This issue, despite its practical significance, has not been thoroughly addressed in our recent work \cite{Zhang2025PRL}. In what follows, we close this crucial gap by establishing three results on the tightness of the Krylov bounds, with proofs deferred to the Supplementary Notes.

We first study how fast $B_n^{(\mathsf{Kry})}$ converges to the QFI as $ n $ increases. To do this, we adopt the following figure of merit
\begin{equation}
    \mathcal{E}_n^{(\mathsf{Kry})} \coloneqq \frac{\abs{B_n^{(\mathsf{Kry})} - F_Q}}{F_Q},
\end{equation}
which is the relative gap between the $n$-th Krylov bound and the true QFI. We are interested in the behavior of $\mathcal{E}_n^{(\mathsf{Kry})}$ as $n$ increases, which is described by the following theorem.
\begin{thm}\label{ThmExpDecay}
    $\mathcal{E}_n^{(\mathsf{Kry})}$ exponentially decreases to zero as $n$ increases,
    \begin{equation}
        \mathcal{E}_n^{(\mathsf{Kry})} \leq 4 \left[ \frac{\sqrt{\kappa(\rho)}-1}{\sqrt{\kappa(\rho)}+1} \right]^{2n}, \label{ExpDecay}
    \end{equation}
    where $ \kappa(\rho) $ is equal to $p_{\textrm{max}}/p_{\textrm{min}}$ for a full-rank $\rho$ and $2p_{\textrm{max}}/p_{\textrm{min}}$ otherwise. Here $p_{\textrm{max}}$ and $p_{\textrm{min}}$ denote the maximal and minimal nonzero eigenvalues of $\rho$, respectively.
\end{thm}
\noindent
We elaborate in Supplementary Note 1 that $\kappa(\rho)$ is the so-called condition number defined as $\kappa=\lambda_{\max}/\lambda_{\min}$ \cite{Liesen2012}, where $\lambda_{\max}$ and $\lambda_{\min}$ are the largest and smallest nonzero eigenvalues of $\frac{1}{2}(\openone\otimes\rho+\rho^T\otimes\openone)$, respectively.
It is expected from Theorem \ref{ThmExpDecay} that the Krylov bounds at low orders already provide reasonably tight approximations to the QFI, owing to the exponential convergence of $\mathcal{E}_n^{(\mathsf{Kry})}$. Moreover, the closer $ \kappa(\rho) $ is to 1, the faster this convergence occurs. Roughly speaking, this means that the relative gap $\mathcal{E}_n^{(\mathsf{Kry})}$ closes more rapidly when $\rho$ is more mixed. This behavior contrasts sharply
with the previous result that polynomial lower bounds may exhibit substantial
gaps from the QFI particularly when $\rho$ is highly mixed.

We next compare the tightness of the Krylov bounds with that of polynomial bounds. Note that the state-of-the-art polynomial bounds are the Taylor bounds \cite{Rath2021PRL,Vitale2024PQ}, given by
\begin{align}\label{defTaylorBounds}
    B_n^{(\mathsf{Tay})} & = 2\tr\Biggl( \sum_{l=0}^n \left(\rho \otimes \openone - \openone \otimes \rho \right)^2
    \left( \openone \otimes \openone - \rho \otimes \openone \right. \nonumber                                      \\
                         & \quad - \left. \openone \otimes \rho \right)^l S(H \otimes H) \Biggr),
\end{align}
where $S$ represents the swap operator and $\openone$ is the identity operator. We denote by $\mathcal{E}_n^{(\mathsf{Tay})}$ the relative gap associated with $B_n^{(\mathsf{Tay})}$, that is,
$\mathcal{E}_n^{(\mathsf{Tay})} \coloneqq \abs{B_n^{(\mathsf{Tay})} - F_Q}\big/{F_Q}$. It has been shown that $B_n^{(\mathsf{Tay})}$ converges exponentially to $F_Q$ as $n\rightarrow\infty$ \cite{Rath2021PRL,Vitale2024PQ}, implying that $\mathcal{E}_n^{(\mathsf{Tay})}$ also decreases exponentially to zero as $n\rightarrow\infty$. So, an interesting question is to compare $\mathcal{E}_n^{(\mathsf{Kry})}$ with $\mathcal{E}_n^{(\mathsf{Tay})}$ in terms of their convergence behaviors, which is addressed in the following theorem.
\begin{thm}\label{ThmKrylovBetter}
    The two relative gaps $ \mathcal{E}_n^{(\mathsf{Kry})} $ and $ \mathcal{E}_{n}^{(\mathsf{Tay})} $ satisfy
    \begin{equation}
        \mathcal{E}_n^{(\mathsf{Kry})} \leq \mathcal{E}_{2n-1}^{(\mathsf{Tay})},
    \end{equation}
    that is, the $n$-th Krylov bound $ B_n^{(\mathsf{Kry})} $ is generally tighter than the $(2n-1)$-th Taylor bound $ B_{2n-1}^{(\mathsf{Tay})} $ for all $ n=1,2,\cdots $.
\end{thm}
\noindent Note that both $ B_n^{(\mathsf{Kry})} $ and $ B_{2n-1}^{(\mathsf{Tay})} $ require evaluating polynomials of $ \rho $ with degree $ 2n+1 $. This implies that the resource demands in estimating these bounds are comparable (a detailed resource comparison can be found in Ref.~\cite{Zhang2025PRL}). We therefore deduce from Theorem \ref{ThmKrylovBetter} that the Krylov bounds can provide more accurate estimates than the Taylor bounds without increasing resource demands. This result highlights the superiority of the KST in estimating the QFI, making it a more effective choice for practical applications in quantum metrology and related fields.

We now examine the condition under which the Krylov bounds can match the QFI exactly at
low orders. Note that the unavoidable gap between polynomial bounds and the QFI introduces
systematic errors in QFI estimation. These systematic errors cannot be mitigated by repeated
measurements, thereby imposing a fundamental limitation on the accuracy of polynomial
approximations. We highlight that the KST opens the possibility of overcoming this fundamental limitation, as the highest order Krylov bound $ B_{n^*}^{(\mathsf{Kry})} $ can match the QFI exactly. The following theorem establishes an upper bound on the value of $ n^* $.
\begin{thm}\label{ThmSmallNStar}
    Let $ S \coloneqq \{p_i + p_j : p_i \neq p_j\}$ consist of the sums of all distinct pairs of eigenvalues of $\rho$. Then we have
    \begin{equation}
        n^* \leq \mathcal{N}[S],
    \end{equation}
    where $\mathcal{N}[S]$ denotes the number of distinct values in the set $ S $.
\end{thm}
\noindent Noting that $\mathcal{N}[S]$ is upper bounded by $r(r+1)/2$ with $r$ being the rank of $\rho$, we have $n^* \leq r(r+1)/2$. This implies that $n^*$ is relatively small whenever $\rho$ is of low rank. It has been acknowledged that states of low ranks arise frequently in quantum information processing tasks \cite{Gross2010,Liu2012}. This is because many protocols require the preparation
of pure states, and in the presence of unavoidable but weak noise, the resulting
mixed states typically retain low rank. We therefore expect from Theorem \ref{ThmSmallNStar} that $n^*$ is small in many relevant scenarios, for which the KST can overcome the fundamental limitation posed by systematic errors and yield an accurate estimate of the QFI with low resource demands. By the way, we present in the Method section a formula for calculating the exact value of $n^*$.

So far, we have presented our analytical findings. We clarify that Theorems \ref{ThmExpDecay} and \ref{ThmSmallNStar} are complementary to each other. Theorem \ref{ThmExpDecay} is particularly useful when $\rho$ is highly mixed and has small $\kappa(\rho)$. For example, for a full-rank $\rho$ with $\kappa(\rho)\leq 11.2$, the third order Krylov bound already provides an estimate of the QFI with a relative error below $0.1$. In contrast, Theorem \ref{ThmSmallNStar} is particularly useful when $\rho$ is nearly pure or of low rank. We would like to add that these two theorems only provide sufficient conditions for efficient and accurate estimation of the QFI with low-order Krylov bounds. It is still possible that the Krylov bounds at low orders can provide tight approximations to the QFI even when $\rho$ is not highly mixed or of low rank. This is because the tightness of the Krylov bounds depends not only on $\rho$ but also on $H$. Below, we conduct extensive numerical simulations to confirm our analytical findings.

\begin{figure}[]
    \centering
    \includegraphics[width=\linewidth]{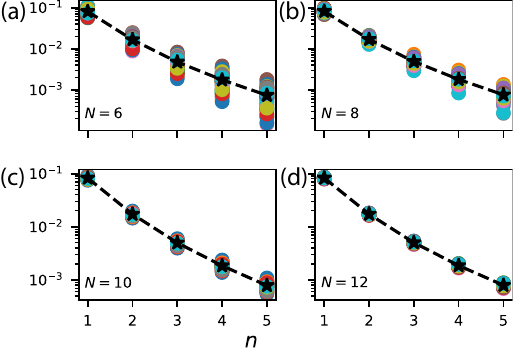}
    \caption{\textbf{Relative gap $ \mathcal{E}_n^{(\mathsf{Kry})} $ as a function of $n$.}  The four subplots correspond to (a) $N=6$, (b) $N=8$, (c) $N=10$, and (d) $N=12$. We generate 100 relative gaps $ \mathcal{E}_n^{(\mathsf{Kry})} $ for a fixed $n$ in each subplot by randomly choosing $100$ full-rank states. The dashed line in each subplot is the average of $100$ relative gaps.}
    \label{Fig2RelativeGap}
\end{figure}

\vspace{1em}
\noindent\textbf{Numerical demonstration}\\
To demonstrate the convergence of the Krylov bounds, we numerically compute the relative gap $\mathcal{E}_n^{(\mathsf{Kry})}$ for different orders $n$. Throughout this section, we fix the operator $H$ as $H =\frac{1}{2} \sum_{i=1}^N \sigma_z^{(i)}$, where $ N $ is the number of qubits and $ \sigma_z^{(i)} $ is the Pauli-$Z$ matrix acting on the $ i $-th qubit. This Hamiltonian has been widely employed in QFI-based criteria for entanglement detection \cite{Boixo2008PRL,Pezze2009PRL,Toth2018PRL,Toth2020PRL,Ren2021PRL,Tan2021PRL,Yang2024PRL}. Figure \ref{Fig2RelativeGap} shows $\mathcal{E}_n^{(\mathsf{Kry})}$ as a function of $n$. The four subplots therein correspond to $N=6,8,10,12$ qubits, respectively. To enhance the statistical relevance of our numerical simulations, we generate 100 relative gaps $\mathcal{E}_n^{(\mathsf{Kry})}$ for a fixed $n$ in each subplot by randomly choosing $100$ full-rank quantum states. We see from Fig.~\ref{Fig2RelativeGap} that all the relative gaps $\mathcal{E}_n^{(\mathsf{Kry})}$ are below $0.1$ when $n\geq 2$, indicating that the Krylov bounds at low orders already provide reasonably tight approximations to the QFI. This observation is consistent with Theorem \ref{ThmExpDecay}, which states that $\mathcal{E}_n^{(\mathsf{Kry})}$ decreases exponentially to zero as $n$ increases. To highlight this convergence, we also compute the average of these 100 relative gaps for each $n$. These averages, shown as dashed lines in the figure, clearly demonstrate the exponential decrease of the relative gap predicted by Theorem~\ref{ThmExpDecay}.

\begin{figure}[]
    \centering
    \includegraphics[width=\linewidth]{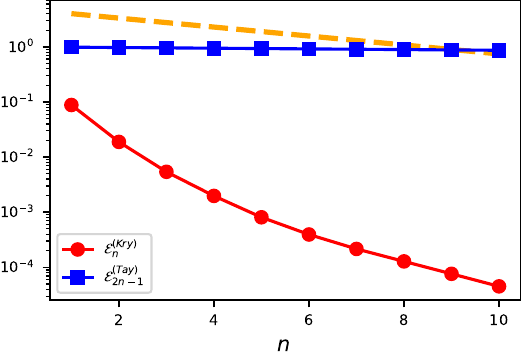}
    \caption{\textbf{Numerical results for $\mathcal{E}_n^{(\mathsf{Kry})}$ and $\mathcal{E}_{2n-1}^{(\mathsf{Tay})}$ as functions of the order $n$.} These two relative gaps are represented by the red solid line with dot markers and the blue solid line with square markers, respectively. The data points in this plot are associated with a randomly generated full-rank quantum state of $N=8$ qubits. The orange dashed line represents the theoretical upper bound on $\mathcal{E}_n^{(\mathsf{Kry})}$  predicted in Theorem \ref{ThmExpDecay}.}
    \label{Fig3FairComparison}
\end{figure}

To compare the tightness of the Krylov bounds with that of the Taylor bounds, we numerically compute both $\mathcal{E}_n^{(\mathsf{Kry})}$ and $\mathcal{E}_{2n-1}^{(\mathsf{Tay})}$ for different $n$. The results are presented in Fig.~\ref{Fig3FairComparison}, where $\mathcal{E}_n^{(\mathsf{Kry})}$ and $\mathcal{E}_{2n-1}^{(\mathsf{Tay})}$ are represented by the red solid line with dot markers and the blue solid line with square markers, respectively. We also show the theoretical upper bound on $\mathcal{E}_n^{(\mathsf{Kry})}$ predicted in Theorem \ref{ThmExpDecay} as a dashed line for comparison. The data points in Fig.~\ref{Fig3FairComparison} are associated with a randomly generated full-rank quantum state of $N=8$ qubits.
We see from Fig.~\ref{Fig3FairComparison} that $\mathcal{E}_n^{(\mathsf{Kry})}$ is considerably smaller than $\mathcal{E}_{2n-1}^{(\mathsf{Tay})}$ for all $n$, in agreement with Theorem \ref{ThmKrylovBetter}. Notably, while both $\mathcal{E}_n^{(\mathsf{Kry})}$ and $\mathcal{E}_{2n-1}^{(\mathsf{Tay})}$ decrease exponentially to zero as $n$ increases, the slope of $\mathcal{E}_n^{(\mathsf{Kry})}$ is evidently steeper than that of $\mathcal{E}_{2n-1}^{(\mathsf{Tay})}$. This indicates that the Krylov bounds converge to the QFI much faster than the Taylor bounds. Further, to exclude the possibility that the results observed in Fig.~\ref{Fig3FairComparison} are merely coincidental, we have conducted extensive numerical experiments by randomly generating full-rank states for different qubits. These experiments consistently yield similar results and reinforce the robustness of our observations.

\begin{figure}[]
    \centering
    \includegraphics[width=\linewidth]{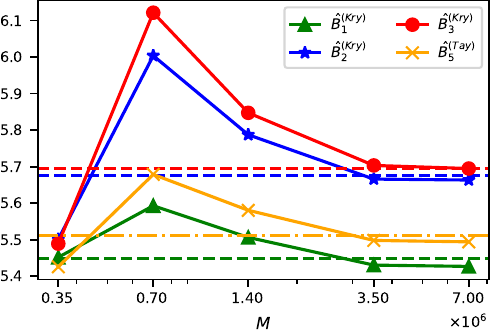}
    \caption{\textbf{Numerical simulations for estimating different bounds.} The estimates of the three Krylov bounds, $\hat{B}_n^{(\mathsf{Kry})}$, $n=1,2,3$, are obtained by numerically simulating the KST for a randomly generated rank-$2$ state of $N=6$ qubits. Different solid lines correspond to different estimates of the bounds, as indicated in the legend. We also show the estimate of the Taylor bound $\hat{B}_5^{(\mathsf{Tay})}$ for comparison. All estimates are plotted as functions of the number $M$ of classical shadows required.   
    The dashed lines correspond to the exact values of the respective bounds.}
    \label{Fig4KSTforRank2}
\end{figure}

To demonstrate the exactness of the Krylov bounds for low-rank states, we numerically simulate the KST to estimate the Krylov bounds $B_n^{(\mathsf{Kry})}$. Here we use the so-called batch shadow method \cite{Rath2023PQ} to accelerate the classical postprocessing in our KST (see the Method section for details).
We denote by $\hat{B}_n^{(\mathsf{Kry})}$ the estimate of $B_n^{(\mathsf{Kry})}$. Figure \ref{Fig4KSTforRank2} shows $\hat{B}_n^{(\mathsf{Kry})}$ as a function of the number $M$ of classical shadows consumed in the KST. Here we randomly generate a rank-2 state of $N=6$ qubits, for which $n^*$ is no more than $3$ according to Theorem \ref{ThmSmallNStar}. We also show the estimate of the Taylor bound $\hat{B}_5^{(\mathsf{Tay})}$ for comparison. As can be seen from Fig.~\ref{Fig4KSTforRank2}, the estimated values of each bound converge to their true values (represented by dashed lines) as $M$ increases. Notably, both
$\hat{B}_2^{(\mathsf{Kry})}$ and $\hat{B}_3^{(\mathsf{Kry})}$ are closer to the QFI (top dashed line) than $\hat{B}_5^{(\mathsf{Tay})}$ for the
same value of $M$ (when $M$ is sufficiently large, e.g., $M \geq 10^6$). Physically speaking, this means that the two estimates of our bounds provide better
approximations to the QFI than the Taylor bound without increasing resource demands. This observation is consistent with Theorem \ref{ThmKrylovBetter}.

\begin{figure}[]
    \centering
    \includegraphics[width=\linewidth]{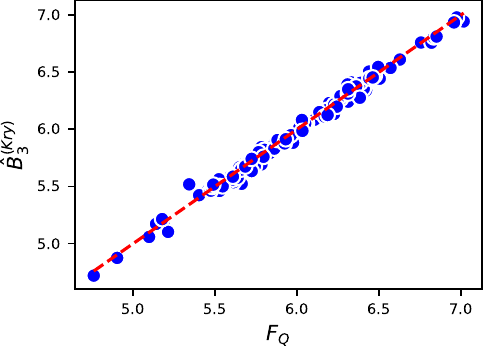}
    \caption{\textbf{Numerical simulations for demonstrating the exact match between the highest Krylov bound and the QFI.} We examine rank-2 states of $N=6$ qubits, for which Theorem~\ref{ThmSmallNStar}
    guarantees that the highest Krylov bound $B_3^{(\mathsf{Kry})}$ can match exactly
    with the QFI. The estimate $\hat{B}_3^{(\mathsf{Kry})}$ is obtained via numerically simulating the KST with
    $M=7\times 10^6$ classical shadows. The plot displays 100 data points, each
    representing a pair $(\hat{B}_3^{(\mathsf{Kry})},F_Q)$ corresponding to a randomly
    generated rank-2 state. The dashed line indicates the diagonal.}
    \label{Fig5hatB3andFQ}
\end{figure}

More importantly, we see from Fig.~\ref{Fig4KSTforRank2} that $\hat{B}_3^{(\mathsf{Kry})}$ ultimately becomes very close to $F_Q$ (represented by the top dashed line), thereby validating the efficiency and accuracy of our KST for low-rank states. To further validate this point, we compute $\hat{B}_3^{(\mathsf{Kry})}$ and $F_Q$ for 100 randomly generated rank-2 states of $N=6$ qubits, using $M=7\times10^6$ classical shadows. The numerical results are presented in Fig. \ref{Fig5hatB3andFQ}, where each point represents a pair of estimates $(\hat{B}_3^{(\mathsf{Kry})}, F_Q)$ for a randomly generated rank-2 state. We see that these points are tightly clustered around the diagonal, indicating a strong correlation between $\hat{B}_3^{(\mathsf{Kry})}$ and $F_Q$. This further validates the effectiveness of our KST in estimating the QFI for low-rank states.

\begin{figure}[]
    \centering
    \includegraphics[width=\linewidth]{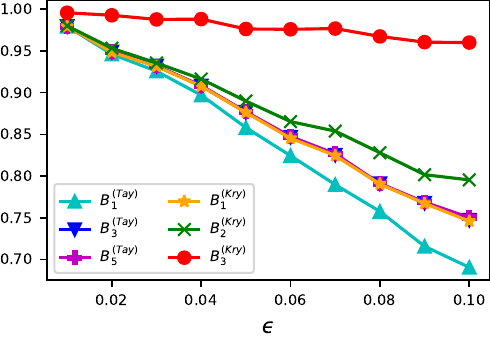}
    \caption{\textbf{Detection ratios as functions of the noise strength $\epsilon$ for different lower bounds.} We randomly generate $10{,}000$ states of the form (\ref{rho-exam}) for $N=6$ qubits and numerically figure out the ratio defined by Eq.~(\ref{ratio}). Here we examine $B_n^{(\mathsf{Kry})}$ and $B_{2n-1}^{(\mathsf{Tay})}$ with $n=1,2,3$, as shown in the legend.}
    \label{Fig6Efficiency}
\end{figure}

Lastly, let us furnish a concrete scenario to demonstrate how the Krylov bounds help to unlock the effectiveness of QFI-based applications. We would like to take entanglement detection as an example. Consider states of the form
\begin{align}\label{rho-exam}
   \rho = (1-\epsilon)\ket{\psi}\!\bra{\psi} + \epsilon\, \sigma^{(\text{noise})},
\end{align}
i.e., the mixture of a pure state and a mixed state stemming from noise. Here $\epsilon$ characterizes the noise strength. When $\sigma^{(\mathrm{noise})}$ is chosen as the maximally mixed state, $\rho$
reduces to the pseudopure state studied in Ref.~\cite{Zhang2025PRL}, corresponding
to a white-noise model. To explore the effectiveness of different lower bounds in entanglement detection, we randomly generate $10{,}000$ states of the form (\ref{rho-exam}) for the $N=6$ qubits and numerically figure out the ratio
\begin{equation}\label{ratio}
    \mathsf{num}(B)/\mathsf{num}(F_Q).
\end{equation}
Here, ${\textrm{num}(F_Q)}$ denotes the number of states that are detected to be entangled by the QFI; that is, $\rho$ is detected to be entangled if $F_Q>N$. Similarly, ${\textrm{num}(B)}$ represents the number of states that are detected to be entangled by the lower bound $B$ in question, that is, $\rho$ is detected to be entangled if $B>N$. Physically speaking, the ratio in Eq.~(\ref{ratio}) represents the effectiveness of $B$ as a lower bound for the QFI in entanglement detection. Apparently, the closer the ratio is to $1$, the more effective the bound $B$ is. The numerical results are shown in Fig.~\ref{Fig6Efficiency}, where the ratio is plotted as a function of $\epsilon$. We see from Fig.~\ref{Fig6Efficiency} that the ratio for $B_n^{(\mathsf{Kry})}$ is significantly higher than that for $B_{2n-1}^{(\mathsf{Tay})}$ across the entire range of $\epsilon$ examined. This implies that $B_n^{(\mathsf{Kry})}$ outperforms $B_{2n-1}^{(\mathsf{Tay})}$ by a significant margin in terms of effectiveness in entanglement detection. Notably, the ratio for $B_3^{(\mathsf{Kry})}$ is above $0.95$ for all values of $\epsilon$. This indicates that $B_3^{(\mathsf{Kry})}$ retains almost the full effectiveness of the QFI in detecting entanglement in this case, which is of particular interest in applications.

\vspace{2em}
\noindent\textbf{\large Discussion}\\
The emergence of the noisy intermediate-scale quantum era \cite{Preskill2018} has brought an 
urgent demand for efficient and accurate estimation of the QFI in large-scale quantum systems. As a fundamental quantity that 
underpins quantum metrology, sensing, and information science, the QFI serves 
as an indispensable tool in diverse applications. However, its highly nonlinear dependence on 
the density matrix makes its experimental estimation notoriously challenging. 
The recently proposed KST offers a promising route toward overcoming this 
challenge by enabling systematically improvable and ultimately accurate estimation 
of the QFI through the non-polynomial Krylov bounds. 

In this work, we have rigorously established the practical applicability of the KST by 
addressing its low-order performance—an essential consideration for near-term 
quantum experiments. Our theoretical analysis yields three main results. First, 
the Krylov bounds converge exponentially fast to the QFI, ensuring that even 
low orders provide tight and reliable estimates (see Theorem \ref{ThmExpDecay}). Second, for comparable resource 
costs, the $n$-th Krylov bound surpasses the $(2n-1)$-th Taylor bound---the state-of-the-art polynomial lower bound known so far---thereby demonstrating the superior 
efficiency of the KST over existing polynomial approaches (see Theorem \ref{ThmKrylovBetter}). Third, we identified 
the class of low-rank quantum states, frequently encountered in practical 
settings, for which certain low-order Krylov bounds match exactly with the QFI (see Theorem \ref{ThmSmallNStar}). This exact match eliminates the systematic errors that fundamentally 
limit polynomial approaches. 

Together, these findings uncover the practical advantages of the KST over existing polynomial approaches, holding promise for fully unlocking the effectiveness of QFI-based applications. The methodology also signifies a broader paradigm shift---from polynomial to non-polynomial 
approaches---for estimating highly nonlinear physical quantities. Beyond QFI estimation, 
this framework may be extended to evaluate other functionals of quantum states, 
such as entanglement monotones \cite{2009Horodecki865,2009Guehne1}, coherence measures \cite{2017Streltsov41003,2017Zhang22323,2018Hu1,2018Zhang170501,Du2021PRA,Du2022SCPM&A}, and geometric quantities \cite{2005Carollo157203,2006Zhu77206,2007Zanardi100603,2007Venuti95701,2019Zhang42104,Zhang2022PRA,Zhou2023E}, many of which exhibit similar high nonlinearity.

Looking ahead, the integration of our KST with advanced experimental techniques could further broaden its impact. For instance, the classical shadows required for our protocol can be efficiently generated via randomized Pauli measurements on qubit systems \cite{Elben2022NRP}, making it particularly suitable for estimating the QFI of multi-qubit quantum states. Furthermore, hybrid classical-quantum post-processing strategies \cite{Xin2020PRA,Kwon2021IToC} and error-mitigated shadow tomography \cite{Aasen2024CP,Jnane2024PQ} could be directly incorporated into the KST framework to enhance its performance on current noisy intermediate-scale quantum devices.


\vspace{2em}
\noindent\textbf{\large Methods}\\
\noindent\textbf{Formula for calculating $ n^* $}\\
The value of $ n^* $ is determined by both the state $ \rho $ and the operator $ H $. While Theorem \ref{ThmSmallNStar} provides a useful upper bound on $ n^* $ solely based on $\rho$, we hereby specify how $H$ affects $n^*$ and present a formula for calculating the exact value of $ n^* $. To this end, we introduce a subset $J$ of $S$ as follows:
\begin{equation}
    J=\{\mu \in S :  \langle i|H|j \rangle = 0~ \text{for all } p_i\neq p_j \text{ with } p_i + p_j = \mu\}.
\end{equation}
Note that, given an element $ \mu \in S $, there may be multiple pairs of eigenvalues $ p_i \neq p_j $ such that $ p_i + p_j = \mu $. The condition for $ \mu $ to belong to $ J $ is that the matrix elements of $ H $ between the corresponding eigenspaces are all zero. We denote by $ \mathcal{N}[J] $ the cardinality of the set $ J $. We then have
\begin{equation}
    n^* = \mathcal{N}[S] - \mathcal{N}[J],
\end{equation}
which gives a formula for calculating the exact value of $ n^* $. The proof of this formula is given in the Supplementary Note 4.


\begin{table}
        \begin{minipage}{\linewidth}
        \rule{\linewidth}{.8pt}\\
        \textbf{Function} \textsc{KrylovBounds}($ n,S(\rho;M) $)
        \\
        \rule{\linewidth}{0.4pt}
        \begin{enumerate}
            \item Import $ S(\rho;M)=\{\hat{\rho}^{(1)},\ldots, \hat{\rho}^{(M)}\} $; \hfill \texttt{// Load classical shadows}
                \item Split $ S(\rho;M) $ into $ (2n+1) $ equally sized subsets $ S_1, \ldots, S_{2n+1} $ and compute $$ \hat{\rho}_i = \frac{1}{\lfloor M/(2n+1)\rfloor} \sum_{j\in S_i} \hat{\rho}^{(j)}, \quad i = 1,\cdots,2n+1;$$ \hfill \texttt{// Construct $ 2n+1 $ batch shadows}
            \item \textbf{For} $ k = 0 $ \textbf{to} $ 2n-1 $ \textbf{do}
                  \begin{align*}
                      \hat{T}_k \leftarrow \frac{1}{ (k+2)! \binom{2n+1}{k+2} } \sum_{i_1 \neq \cdots \neq i_{k+2}} \\ \operatorname{tr} \left( i[\hat{\rho}_{i_1}, H] \mathcal{R}_{\hat{\rho}_{i_2}} \circ \cdots \circ \mathcal{R}_{\hat{\rho}_{i_{k+1}}}(i[\hat{\rho}_{i_{k+2}}, H])  \right); 
                    \end{align*} \hfill \texttt{// Estimate $ T_k $ using U-statistics}
            \item $ A \leftarrow [\hat{T}_{i+j-1}]_{i,j=1}^{n}$; $b \leftarrow [\hat{T}_{i-1}]_{i=1}^{n} $; \hfill \texttt{//  Compute $ A $ and $ b $}
            \item \textbf{Return} $ \hat{B}_n^{(\mathsf{Kry})} \leftarrow b^TA^{-1}b $. \hfill \texttt{//  Estimate $ B_n^{(\mathsf{Kry})} $ via Eq.~(\ref{main-finding-2nd})}
        \end{enumerate}
        \rule{\linewidth}{0.4pt}
    \end{minipage}
    \caption{\label{alg:KST} \textbf{Procedure for estimating the $n$-th Krylov bound.}}
\end{table}


\vspace{1em}
\noindent\textbf{Procedure for estimating $B_n^{(\mathsf{Kry})}$}\\
Let us recapitulate that $B_n^{(\mathsf{Kry})}$ can be expressed as follows \cite{Zhang2025PRL}:
\begin{eqnarray}\label{main-finding-2nd}
    B_n^{(\mathsf{Kry})}=b^TA^{-1}b,
\end{eqnarray}
where $A$ is an $n\times n$ matrix with its $(i,j)$-th entry given by $T_{i+j-1}$ and $b$ is an $n$-dimensional vector with its $i$-th entry given by $T_{i-1}$. Here $T_k$ is a polynomial of $\rho$, defined as 
\begin{eqnarray}
    T_k\coloneqq\tr\left(i[\rho,H]\mathcal{R}_\rho^{k}(i[\rho,H])\right).
\end{eqnarray}
Therefore, to obtain $B_n^{(\mathsf{Kry})}$, we need to estimate all the $T_k$ for $k=0,\cdots,2n-1$. As detailed in our recent work \cite{Zhang2025PRL}, this task can be accomplished by
employing shadow tomography, which enables the estimation of polynomial functions of
the density matrix $\rho$ through U-statistics \cite{Huang2020}. However, the direct use of
U-statistics is computationally expensive. We here use the batch shadow method \cite{Rath2023PQ} to accelerate the classical postprocessing in our KST. That is, we first construct $(2n+1)$ batch shadows from the original set of classical shadows and then apply U-statistics to these batch shadows to obtain the estimates for all $T_k$. The detailed procedure can be found in Tab.~\ref{alg:KST}.

\vspace{2em}
\noindent\textbf{Data availability}\\
The datasets generated and/or analyzed during the current study are not publicly available due to the large volume of raw numerical data but are available from the corresponding author on reasonable request. The present study did not use any publicly available sequence data.

\vspace{2em}
\noindent\textbf{Code availability}\\
The codes used for this study are available from Y.-H.W. upon
request.


%



\vspace{2em}
\noindent\textbf{Acknowledgements}\\
This work is supported by the National Natural Science Foundation of China under Grant No.~12275155 and the Shandong Provincial Young Scientists Fund under Grant No.~ZR2025QB16. The funders played no role in study design, data collection, analysis and interpretation of data, or the writing of this manuscript. The scientific calculations in this paper have been done on the HPC Cloud Platform of Shandong University.

\vspace{2em}
\noindent\textbf{Author contributions}\\
D.-J.Z. conceived the idea, outlined the theoretical framework, and supervised the project. Y.-H.W. proved the theorems and conducted the numerical simulations. All authors contributed to the preparation of the manuscript.

\vspace{2em}
\noindent\textbf{Competing interests}\\
The authors declare no competing interests.



\section*{Supplementary material}

# Supplemental Information for "Superiority of Krylov shadow tomography in estimating quantum Fisher information: From bounds to exactness"

Yuan-Hao Wang and Da-Jian Zhang (corresponding author)[*]
*Department of Physics, Shandong University, Jinan 250100, China*
(Dated: February 13, 2026)

[*] [zdj@sdu.edu.cn](mailto:zdj@sdu.edu.cn)

## SUPPLEMENTARY NOTE 1: PROOF OF THEOREM 1

To prove Theorem 1, we recall that $B_n^{(\text{Kry})} := \|L_n\|_\rho^2$ where $L_n \in \mathcal{K}_n$ is chosen to be as close as possible to $L$. We have shown [1] that

$$F_Q - B_n^{(\text{Kry})} = \|L - L_n\|_\rho^2. \tag{S1}$$

Let $|X\rangle\rangle$ denote the vectorization of $X$, obtained by stacking the columns of $X$ into a single column vector. That is,

$$|X\rangle\rangle = \sum_{i,j} X_{ij} |i\rangle \otimes |j\rangle, \tag{S2}$$

with $X_{ij}$ denoting the elements of the matrix $X$. Noting that $\text{tr}\left[\rho \frac{XY+YX}{2}\right] = \text{tr}\left[X\mathcal{R}_\rho(Y)\right]$ and using the formula $|AXB\rangle\rangle = A \otimes B^T |X\rangle\rangle$ for two matrices $A$ and $B$ [2], where $T$ denotes the transpose, we can rewrite the induced norm as

$$\|X\|_\rho^2 = \langle\langle X| \left[\frac{1}{2}(\mathbb{1} \otimes \rho + \rho^T \otimes \mathbb{1})\right] |X\rangle\rangle. \tag{S3}$$

We see from Eq. (S3) that the induced norm can be interpreted as the vector norm adopted in the Conjugate Gradient method [3]. Using the known result on the error in the Conjugate Gradient method (see Corollary 5.6.7 in Ref. [4]), we obtain

$$\frac{\|L - L_n\|_\rho}{\|L\|_\rho} \leq 2 \left(\frac{\sqrt{\kappa} - 1}{\sqrt{\kappa} + 1}\right)^n. \tag{S4}$$

Here, $\kappa$ is known as the condition number, defined as $\kappa = \lambda_{\max}/\lambda_{\min}$, where $\lambda_{\max}$ and $\lambda_{\min}$ are respectively the largest and smallest nonzero eigenvalues of the matrix $\frac{1}{2}(\mathbb{1} \otimes \rho + \rho^T \otimes \mathbb{1})$. Let $p_{\max}$ and $p_{\min}$ denote the maximal and minimal nonzero eigenvalues of $\rho$, respectively. It is easy to see that $\kappa$ is equal to $p_{\max}/p_{\min}$ for a full-rank $\rho$ and $2p_{\max}/p_{\min}$ otherwise. Note that $F_Q = \|L\|_\rho^2$ [1]. The proof of Theorem 1 is completed by inserting $F_Q = \|L\|_\rho^2$ and Eq. (S1) into Eq. (S4).

## SUPPLEMENTARY NOTE 2: PROOF OF THEOREM 2

To prove Theorem 2, we first need to rewrite the $(2n-1)$-th Taylor bound in a more convenient form. Let us recall that

$$B_{2n-1}^{(\text{Tay})} = 2\,\text{tr}\left(\sum_{l=0}^{2n-1} (\rho \otimes \mathbb{1} - \mathbb{1} \otimes \rho)^2 (\mathbb{1} \otimes \mathbb{1} - \rho \otimes \mathbb{1} - \mathbb{1} \otimes \rho)^l S(H \otimes H)\right), \tag{S5}$$

where $S$ is the swap operator and $\mathbb{1}$ is the identity operator. Noting that $|\mathcal{R}_\rho(X)\rangle\rangle = \frac{1}{2}(\mathbb{1} \otimes \rho + \rho^T \otimes \mathbb{1})|X\rangle\rangle$ and $(|H\rangle\rangle\langle\langle H|)^{T_2} = S(H \otimes H)$, where $T_2$ is the partial transpose with respect to the second subsystem, we obtain, after some algebra,

$$B_{2n-1}^{(\text{Tay})} = \text{tr}\left(i[\rho, H] \sum_{l=0}^{2n-1} 2\,(\text{id} - 2\mathcal{R}_\rho)^l \,(i[\rho, H])\right), \tag{S6}$$

where id denotes the identity superoperator. Besides, using the formula for the sum of a geometric series, we have

$$\sum_{l=0}^{2n-1} 2\,(\text{id} - 2\mathcal{R}_\rho)^l = \mathcal{R}_\rho^{-1}\left(\text{id} - (\text{id} - 2\mathcal{R}_\rho)^{2n}\right). \tag{S7}$$

Inserting Eq. (S7) into Eq. (S6) and noting that $i[\rho, H] = \mathcal{R}_\rho(L)$ and $\text{tr}[\mathcal{R}_\rho(X)Y] = \text{tr}[X\mathcal{R}_\rho(Y)]$ for two matrices $X$ and $Y$, we obtain

$$B_{2n-1}^{(\text{Tay})} = \text{tr}\left[L\left(\text{id} - (\text{id} - 2\mathcal{R}_\rho)^{2n}\right)\mathcal{R}_\rho(L)\right]. \tag{S8}$$

Further, noting that $\|X\|_\rho^2 = \text{tr}[\mathcal{R}_\rho(X)X] = \text{tr}[X\mathcal{R}_\rho(X)]$ for any matrix $X$, we deduce from Eq. (S8) the convenient form

$$B_{2n-1}^{(\text{Tay})} = \|L\|_\rho^2 - \|(\text{id} - 2\mathcal{R}_\rho)^n(L)\|_\rho^2. \tag{S9}$$

Here we have used the fact that $(\text{id} - 2\mathcal{R}_\rho)$ commutes with $\mathcal{R}_\rho$. Recall that $F_Q = \|L\|_\rho^2$ [1]. We deduce from Eq. (S9) that

$$F_Q - B_{2n-1}^{(\text{Tay})} = \|(\text{id} - 2\mathcal{R}_\rho)^n(L)\|_\rho^2. \tag{S10}$$

On the other hand, recall that $L_n \in \mathcal{K}_n$ is chosen to be as close as possible to $L$, that is, $L_n$ attains the following minimum

$$\|L - L_n\|_\rho^2 = \min_{X \in \mathcal{K}_n} \|L - X\|_\rho^2. \tag{S11}$$

Hence,

$$F_Q - B_n^{(\text{Kry})} = \min_{X \in \mathcal{K}_n} \|L - X\|_\rho^2, \tag{S12}$$

which follows by inserting Eq. (S11) into Eq. (S1). Without loss of generality, we can express $X \in \mathcal{K}_n$ as $X = \sum_{k=0}^{n-1} \alpha_k \mathcal{R}_\rho^k(i[\rho, H])$ for some $\alpha_k \in \mathbb{R}$. Noting that $i[\rho, H] = \mathcal{R}_\rho(L)$, we further have that $X = \sum_{k=1}^{n} \alpha_k \mathcal{R}_\rho^k(L)$. Therefore,

$$F_Q - B_n^{(\text{Kry})} = \min_{\substack{\varphi(0)=1 \\ \deg \varphi \leq n}} \|\varphi(\mathcal{R}_\rho)(L)\|_\rho^2, \tag{S13}$$

where $\varphi$ represents a polynomial and $\deg \varphi \leq n$ means that the degree of the polynomial is no more than $n$. Here $\varphi(0) = 1$ means that the constant term in this polynomial is 1. Note that $(\text{id} - 2\mathcal{R}_\rho)^n$ appearing in Eq. (S10) is a polynomial satisfying $\varphi(0) = 1$ and $\deg \varphi \leq n$. Using this fact and comparing Eq. (S10) with Eq. (S13), we conclude that

$$F_Q - B_n^{(\text{Kry})} \leq F_Q - B_{2n-1}^{(\text{Tay})}. \tag{S14}$$

The proof of Theorem 2 is completed by noting that both $F_Q - B_n^{(\text{Kry})}$ and $F_Q - B_{2n-1}^{(\text{Tay})}$ are nonnegative and Eq. (S14) implies $\mathcal{E}_n^{(\text{Kry})} \leq \mathcal{E}_{2n-1}^{(\text{Tay})}$.

## SUPPLEMENTARY NOTE 3: PROOF OF THEOREM 3

To prove Theorem 3, we introduce a sequence of vectors $|\phi_k\rangle$ defined to be the vectorization of $\mathcal{R}_\rho^k(i[\rho, H])$, namely,

$$|\phi_k\rangle = |\mathcal{R}_\rho^k(i[\rho, H])\rangle\rangle, \quad k = 0, 1, 2, \cdots, d^2 - 1, \tag{S15}$$

where $d$ denotes the dimension of the Hilbert space in question. It is not difficult to see that $n^*$ is equal to the rank of the matrix

$$\Phi = [|\phi_0\rangle, |\phi_1\rangle, \cdots, |\phi_{d^2-1}\rangle], \tag{S16}$$

defined by stacking the vectors $|\phi_k\rangle$ as columns. Noting that $\mathcal{R}_\rho(X) = \sum_{kl} \frac{p_k + p_l}{2} X_{kl} |k\rangle\langle l|$ where $X_{kl} = \langle k|X|l\rangle$, we have

$$\mathcal{R}_\rho^k(X) = \sum_{kl} \left(\frac{p_k + p_l}{2}\right)^k X_{kl} |k\rangle\langle l|. \tag{S17}$$

It follows from Eq. (S17) that $\mathcal{R}_\rho^k(i[\rho, H]) = i \sum_{kl} \left(\frac{p_k + p_l}{2}\right)^k (p_k - p_l) H_{kl} |k\rangle\langle l|$, where $H_{kl} = \langle k|H|l\rangle$. We can express $|\phi_k\rangle$ as

$$|\phi_k\rangle = i \sum_{kl} \left(\frac{p_k + p_l}{2}\right)^k (p_k - p_l) H_{kl} |k\rangle \otimes |l\rangle. \tag{S18}$$

4Hence,

$$\Phi = \begin{bmatrix} 0 & 0 & 0 & \cdots \\ i\left(\frac{p_1+p_2}{2}\right)^0 (p_1-p_2)H_{12} & i\left(\frac{p_1+p_2}{2}\right)^1 (p_1-p_2)H_{12} & i\left(\frac{p_1+p_2}{2}\right)^2 (p_1-p_2)H_{12} & \cdots \\ i\left(\frac{p_1+p_3}{2}\right)^0 (p_1-p_3)H_{13} & i\left(\frac{p_1+p_3}{2}\right)^1 (p_1-p_3)H_{13} & i\left(\frac{p_1+p_3}{2}\right)^2 (p_1-p_3)H_{13} & \cdots \\ \vdots & \vdots & \vdots & \ddots \end{bmatrix}. \quad (S19)$$

Note that multiplying a row of a matrix by a nonzero scalar does not change the rank of the matrix. Note also that deleting a row of all zeros does not change the rank of the matrix, too. We can bound the rank of $\Phi$ from above as follows:

$$\text{rank}(\Phi) \leq \text{rank} \begin{bmatrix} 1 & \left(\frac{p_1+p_2}{2}\right)^1 & \left(\frac{p_1+p_2}{2}\right)^2 & \cdots \\ 1 & \left(\frac{p_1+p_3}{2}\right)^1 & \left(\frac{p_1+p_3}{2}\right)^2 & \cdots \\ 1 & \left(\frac{p_1+p_4}{2}\right)^1 & \left(\frac{p_1+p_4}{2}\right)^2 & \cdots \\ \vdots & \vdots & \vdots & \ddots \end{bmatrix}. \quad (S20)$$

The matrix appearing in the right-hand side of Eq. (S20) is a Vandermonde matrix. We clarify that this matrix only consists of the elements $\left(\frac{p_i+p_j}{2}\right)^k$ with $p_i \neq p_j$, because the row of $\Phi$ corresponding to $p_i = p_j$ is zero. The rank of this Vandermonde matrix is therefore no more than $\mathcal{N}[S]$ defined in the main text. This implies that $n^* = \text{rank}(\Phi) \leq \mathcal{N}[S]$.

### SUPPLEMENTARY NOTE 4: PROOF OF THE FORMULA

It is easy to prove the formula. Let $\mu \in J$, that is, for all pairs $p_i \neq p_j$ such that $p_i + p_j = \mu$, there is $H_{ij} = 0$. Note that the rows of $\Phi$ corresponding to these pairs are all zero. So, deleting these rows does not change the rank of $\Phi$. Therefore, whenever we encounter such a $\mu$, we should minus one from $\mathcal{N}[S]$ in order to only count the rows of $\Phi$ that contribute to the rank of $\Phi$. This implies that the rank of $\Phi$ is $\mathcal{N}[S] - \mathcal{N}[J]$. Then the formula follows from $n^* = \text{rank}(\Phi)$.

---


\end{document}